# Laguerre-Gaussian transform for rotating image processing

Dan Wei[1,*], Jiantao Ma[1,*], Tianxin Wang[1], Chuan Xu[1], Yin Cai[1], Lidan Zhang[1], Xinyuan Fang[1], Dunzhao Wei[1], Shining Zhu[1,2], Yong Zhang[1,2,†], Min Xiao[1,2,3]

[1]National Laboratory of Solid State Microstructures, College of Engineering and Applied Sciences, and School of Physics, Nanjing University, Nanjing 210093, China
[2]Collaborative Innovation Center of Advanced Microstructures, Nanjing University, Nanjing 210093, China
[3]Department of Physics, University of Arkansas, Fayetteville, Arkansas 72701, USA
*These authors contribute equally to this work.
†To whom correspondence should be addressed: zhangyong@nju.edu.cn

**Rotation is a common motional form in nature, existing from atoms and molecules, industrial turbines to astronomical objects. However, it still lacks an efficient and reliable method for real-time image processing of a fast-rotating object. Since the Fourier spectrum of a rotating object changes rapidly, the traditional Fourier transform (FT) techniques become extremely complicated and time consuming. Here, we propose a Laguerre-Gaussian (LG) transform to analyze the rotating object with LG-mode[5-7] basis. The rotation operation provides a feasible way to acquire LG spectrum, which is similar to the function of lens in FT. Particularly, the obtained LG spectrum does not change even the object working at a high rotating speed. By analyzing the LG spectrum, one can perform image processing such as reconstruction, edge enhancement, and pattern replication. Such LG transform provides an efficient and convenient way to real-time monitor and analyze a fast-rotating object in scientific research and industry.**

FT decomposes a temporal signal into its constituent frequencies. Because many linear operations including differentiation and convolution are much easier to perform in the frequency domain, FT has been widely applied in analysis of differential equations, FT spectroscopy, and signal processing [1]. In Fourier optics, FT from an optical pattern to its spatial frequency components can be readily achieved by using an optical lens [2]. Based on such optical FT, powerful tools, such as image reconstruction, edge recognition, spatial filtering, computer generated hologram, and image compression [3], have been developed for image processing schemes. However, in real-time monitoring of a rotating object (such as a biological molecule, industrial centrifuge and turbine, and an astronomical object), the typical optical FT techniques wouldn't work as efficiently as usual. Since its Fourier spatial frequency spectrum is changing along with the rotating object, FT-based image processing schemes become complicated and time consuming. In this Article, we propose and experimentally demonstrate to use a Laguerre-Gaussian (LG) transform of a rotating object to overcome this problem. The obtained LG spectrum does not vary with the rotation, which provides great convenience and a nature way for further image processing of a fast-rotating object.

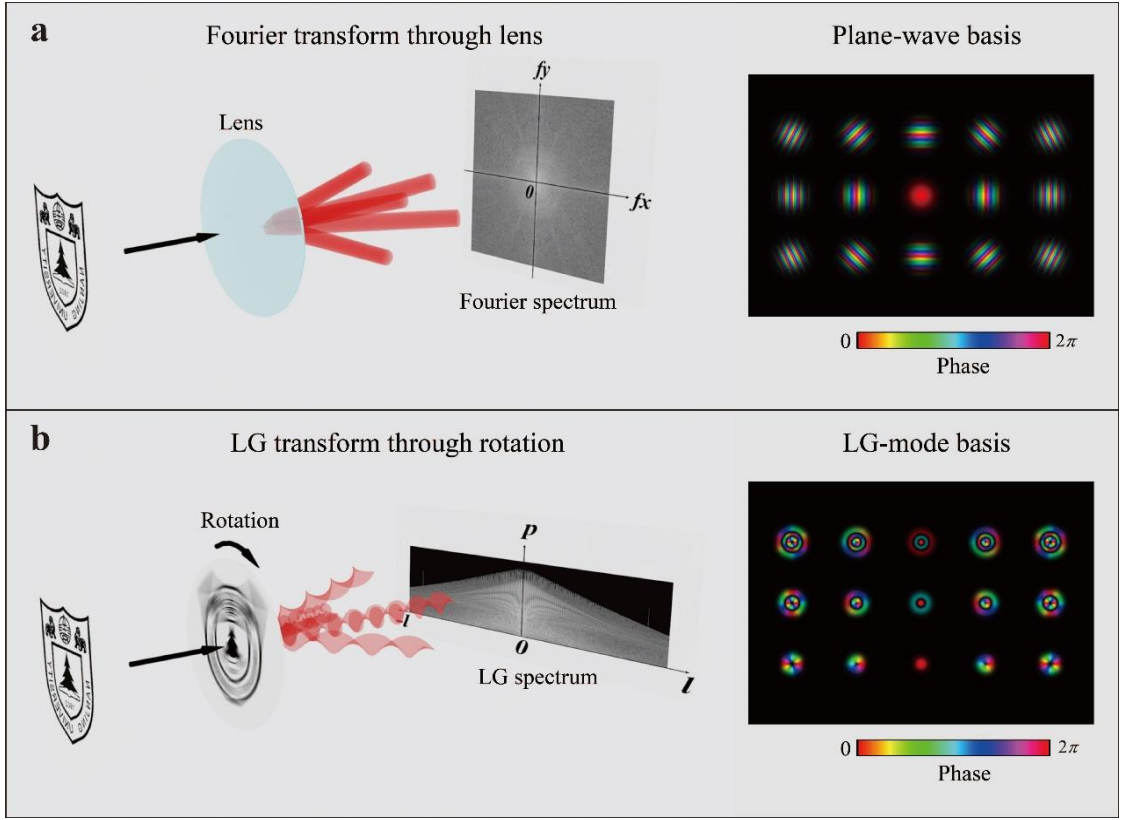

**Fig. 1| Comparison of FT and LG transforms. a**, In FT transform, a spatial image is decomposed into its spatial frequency spectrum by use of an optical lens. However, if the image is rotating, the FT spectrum is changing. **b**, In LG transform, an image is considered as a superposition of various LG modes. Rotating the image is an effective operation to obtain its LG spectrum, which does not change even the object has a fast rotary speed. See Supplementary Section 1 and Fig. S1 for details.

Figure 1 compares the FT and LG transforms of a spatial image. The fundamental difference is the basis used to decompose the target spatial pattern. In FT, the optical image of an object is considered as a superposition of plane waves and the aim is to convert the image into the spatial frequency spectrum. In LG transform, the basis to decompose an optical image is chosen to be the LG modes. As a complete and orthogonal basis, LG modes are also capable to fully represent the spatial structure of a transverse field [4, 5]. LG modes are the Eigen solutions of the paraxial Helmholtz equation [6], which are characterized by an azimuthal index $l$ and a radial index $p$,

$$LG_{p,l}(r,\phi) = \sqrt{\frac{2p!}{\pi(p+|l|)!}} \frac{1}{\omega_0} \left(\sqrt{2}\frac{r}{\omega_0}\right)^{|l|} L_p^{|l|}\left(2\frac{r^2}{\omega_0^2}\right) \exp\left(-\frac{r^2}{\omega_0^2}\right) \exp(il\phi) \\ = LG_{p,l}(r)\exp(il\phi)$$ (1)

Here, $L_p^{|l|}$ are the Laguerre polynomials, $r$ is the radial cylindrical coordinate, $\phi$ is the azimuthal angle, and $\omega_0$ is the beam waist. $LG_{p,l}(r)$ is the radial term of an LG mode. Various methods have been proposed to generate LG modes, such as spiral phase plates[7], spatial light modulators, Q-plates[8], and cavities[9-11]. Because the LG basis naturally features circular symmetry, rotating the object does not change the measured LG spectrum in such case, even when the object rotates at a high speed. Such unique advantage makes the basis of LG modes an ideal candidate to analyze a rotating object.

In addition, rotation is a feasible operation to obtain LG spectrum, which is similar to the function of a lens in optical FT. The azimuthal index $l$ of an LG mode is associated with the orbital angular momentum of light [12], which has been widely investigated in optical communications [13, 14], gravitational-wave detection [15], and nonlinear and quantum optics [16, 17]. Based on the rotational Doppler effect, i.e., the Doppler frequency shift from a spinning object being proportional to $l$ [18-25], one can have the distribution of $l$ index [23, 25]. The $l$ spectrum reflects the azimuthal structure of the object, which can be used to measure the rotary speed [20, 23], azimuthal angle [26] and azimuthal symmetry [4, 5, 22, 23, 26, 27]. However, this is not enough to reconstruct a full image because of the loss of radial information. To fix this problem, we develop an effective method to obtain both azimuthal index $l$ and radial index $p$ for constructing the complete LG spectrum.

Assume that the target rotating field at t = 0 is described as. By using the LG-mode basis, $E(r,\phi)$ can be written as

$$E(r,\phi) = \sum_{l}\sum_{p} A_{p,l} LG_{p,l}(r) \exp(il\phi)$$
$$= \sum_{l} b_{l}(r) \exp(il\phi) \qquad (2)$$

where $A_{p,l}$ is the complex LG spectrum of $E(r,\phi)$ with $\sum_{l}\sum_{p}|A_{p,l}|^2 = 1$. Here, we define

$$b_{l}(r) = \sum_{p} A_{p,l} LG_{p,l}(r), \qquad (3)$$

which represents the overall contributions from the LG modes with the same $l$ but different $p$. $b_{l}(r)$ generally varies with radius $r$ in a complex image.

When the object rotates at an angular frequency $\Omega$, each LG component has a frequency shift $l\Omega$ according to the rotational Doppler effect [18-25]. Then, the rotating field can be written as

$$E(r,\phi,t) = \sum_{l} b_{l}(r) \exp(il\phi) \exp(il\Omega t), \qquad (4)$$

where $t$ denotes time. One can measure $E(r,\phi_0,t)$ by setting a 1D detector array or scanning a single-point detector (SPD) along a certain azimuthal angle $\phi_0$. Through time Fourier transform, $b_{l}(r)$ is calculated from

$$b_{l}(r) = \frac{\exp(-il\phi_0)}{T}\int_{0}^{T} E(r,\phi_0,t)\exp(-il\Omega t)dt, \qquad (5)$$

where $T = \frac{2\pi}{\Omega}$ is the rotation period. Then, the LG spectrum (i.e., $A_{p,l}$) can be deduced according to Eqs. (3) and (5). Here, we develop an efficient algorithm to calculate $A_{p,l}$ by using a sampling number as few as possible without losing the major information of the rotating object. In principle, the minimal sampling points are $2l+2$ and $p+1$ for azimuthal and radial directions, respectively, to obtain all $A_{p,l}$. Note that the fitting parameters such as beam waist and truncation mode number of LG mode are optimized. See Supplementary Sections 2 and 3 for the detailed algorithm.

In the experiment, the rotating pattern is generated by a computer-controlled digital micromirror device (DMD) (Fig. 2a). The rotating period is set to be 50 ms. After using a 4f imaging system to amplify the pattern, a rotating image (~1000 dpi) is detected by a 1D detector array. See Figs. S2 and S3 in Supplementary Information for details of the experimental setup. Figure 2 depicts the reconstruction of a rotating emblem of Nanjing University. See Movie S1 in Supplementary Information. Figures 2b and 2c show the amplitude and phase of the measured LG spectrum. The used LG basis includes about $10^5$ LG components with $l$ ranging from -300 to 300 and $p$ from 0 to 180. Figures 2h-k show the reconstructed images using 1/16, 1/4, 9/16 and the entirely measured LG spectrum, respectively. As shown in Fig. 2i, the image can be well reproduced using 1/4 of the measured LG spectrum because most of the image information are provided by the relatively lower-order LG components. If enlarging the scale of the selected LG modes (Figs. 2f and 2g), the image quality can be further increased (Figs. 2j and 2k).

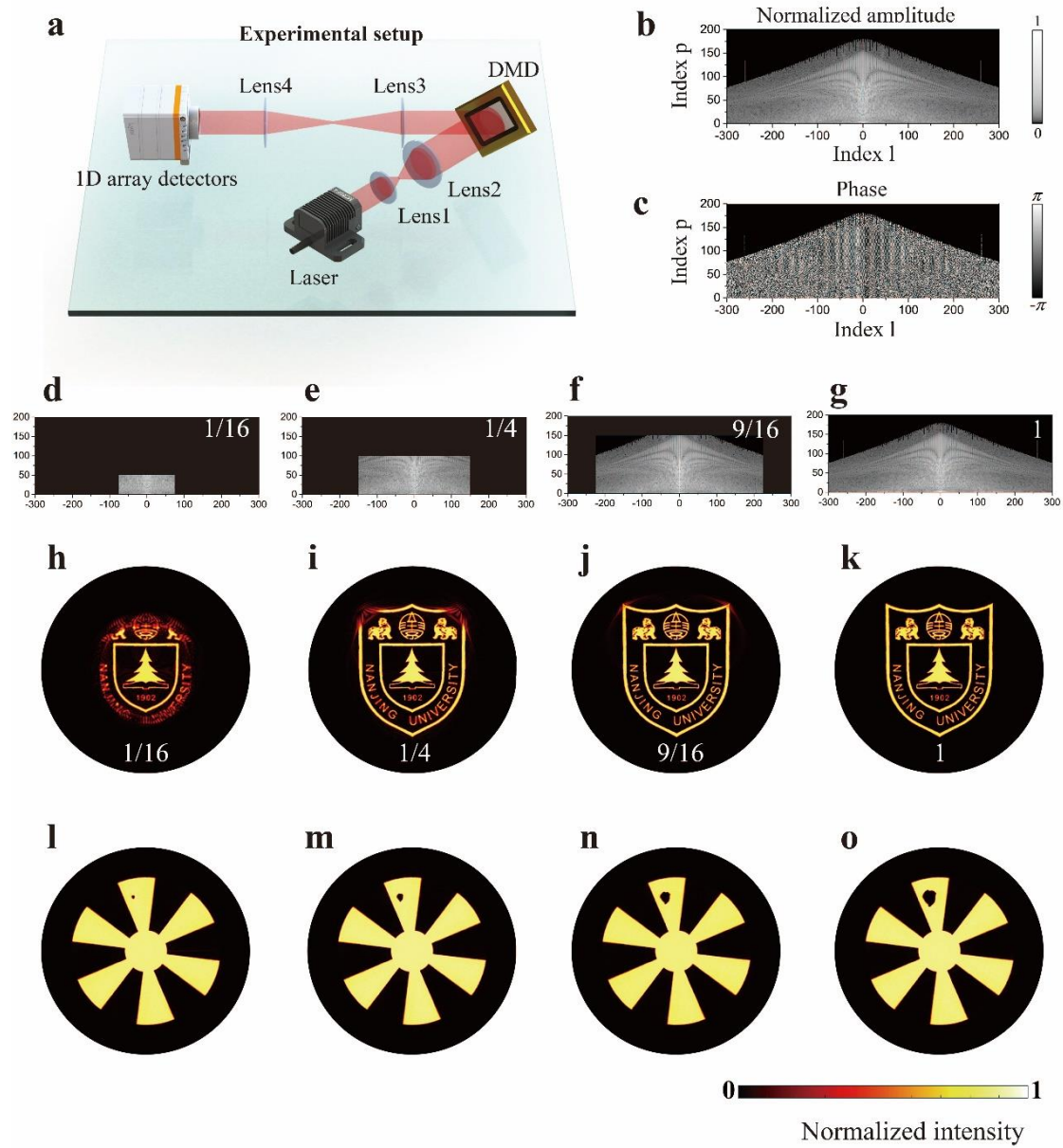

**Fig. 2 | The experimental results. a**, shows the experimental setup. A series of amplitude patterns are playing on the DMD to mimic a rotating object. The rotating field is detected by a 1D array detector, which is set at an azimuthal angle $\phi_0$. **b** and **c** are the amplitude and phase of the measured LG spectrum from a rotating emblem of Nanjing University. **d-g** show used portions of the LG spectrum to reconstruct the emblems displayed in **h-k**, respectively. **l-o** exhibit an example of monitoring the defect evolution in a working fan. The reconstructed frames clearly show a growing hole. See Movies S1 and S2 in Supplementary Information.

Considering that the information captured within one rotation period is enough to reconstruct a full image, this technique can be used for certain practical task, i.e., real-time monitoring the dynamic evolution of a rotating pattern. For example, we track a hole defect in a working fan. Figures 2l-o present four continuous frames of the fan reconstructed from the real-time LG spectra. One can clearly see the growing process of the hole defect when the fan is still at work (see Movie S2 in Supplementary Information). Interestingly, faster the fan rotates, more frames per second can be achieved in principle. Such unique function of LG transform is very helpful for the in-situ monitoring of the status of a rotating object when it is inconvenient or too expensive to completely stop running it for inspection.

Figure 3 shows several typical image processing schemes by selectively modulating the LG spectrum. Figures 3a-e demonstrate the edge enhancement of a fan by performing differential operations on the LG spectrum. The azimuthal and radial edge enhancements by choosing $\phi$ and $r$ as the differential variables can be performed separately (Figs. 3c and 3d) or simultaneously (Fig. 3e). Figures 3f-3j present the symmetric operations of a zebra image. By simply adding a phase $l\Delta\phi$ into each LG component, one can rotate the original image by an angle of $\Delta\phi$ (Fig. 3H). Also, one can flip the image by conjugating the spiral phase of each LG mode (Fig. 3i). By combining above two operations, the image can be folded along $-\Delta\phi/2$ angle (Fig. 3j). Figures 3k-3o depict how to replicate a pattern by manipulating the LG spectrum. The original image is a single leaf (Fig. 3k). Through picking out the LG modes with their $l$ indices being the multiple of 2, 3, or 4, one can obtain double (Fig. 3m), triple (Fig. 3n) or quadruple leaves (Fig. 3o), respectively (see Methods for details).

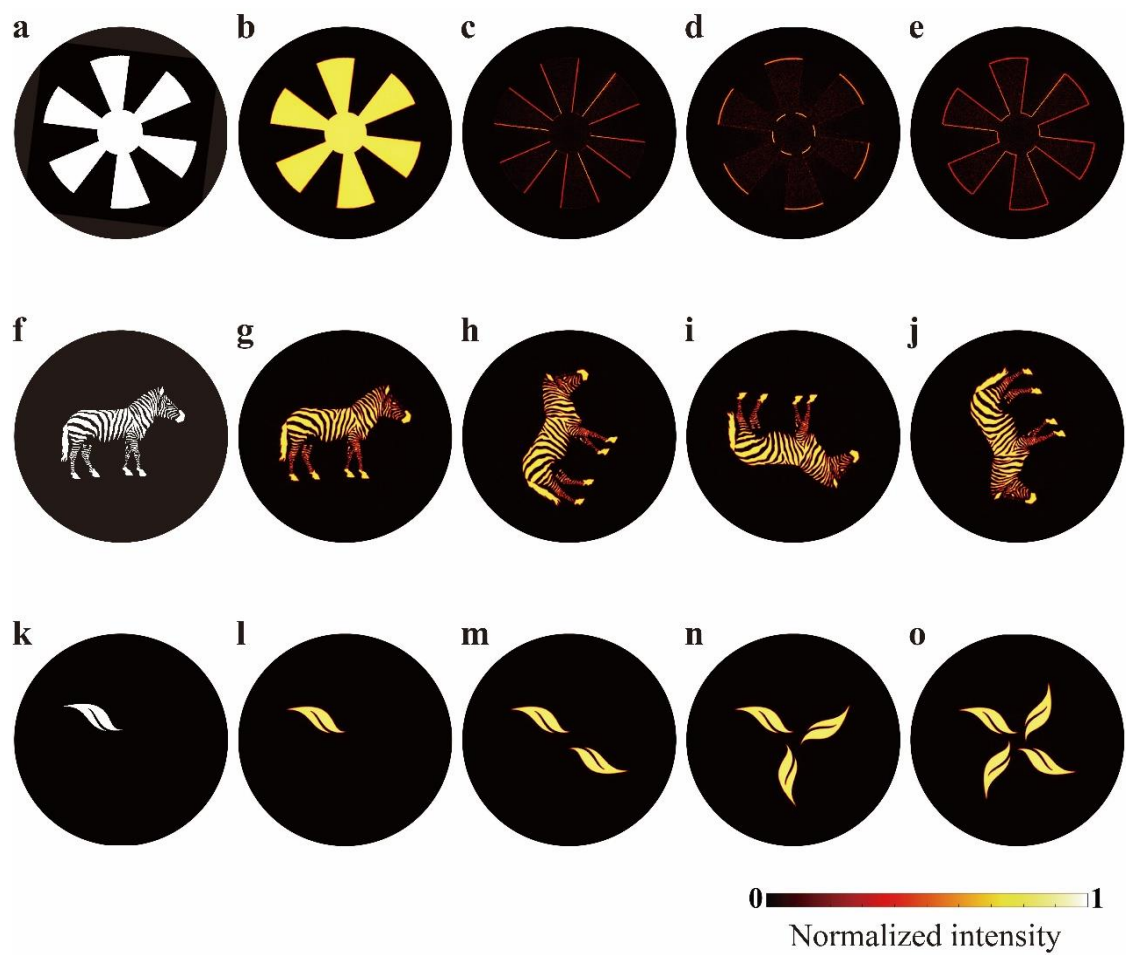

**Fig. 3 | Image processing by manipulating the LG spectrum.** When the original objects, i.e., a fan **a**, a zebra **f**, and a leaf **k**, are rotating, we can well reconstruct their images in **b**, **g**, and **l** by using the LG transform method. Azimuthal and radial edge enhancements in **c-e** can be achieved by choosing either $\phi$ **c**, $r$ **d**, or both **e** as differential variables. Symmetric operations can be performed by adding a phase $l\Delta\phi$ into the LG spectrum **h**, conjugating the phase of the LG spectrum **i**, or combining them together **j**. Pattern replication can be realized by selecting the LG components with their $l$ indices being the multiple of 2 **m**, 3 **n**, or 4 **o**.

In addition, our method can be performed by using one SPD, which, in comparison to array detectors, has the advantages of faster response time and broader frequency bandwidth, and therefore, is capable to track higher-speed rotating object. The schematic experimental setup is shown in Fig. 4a. A chopper covered by the numbers from 1 to 5 serves as the rotating object. It works at 100 Hz frequency. The rotating optical field is detected by a 1-MHz SPD, which scans along the radial direction. The reconstructed image using the LG spectrum is shown in Fig. 4b, which well reproduces the original five numbers. In theory, the maximal measurable rotation speed of our system is up to 2,000 revolutions per second, which can be further enhanced by increasing the bandwidth of the used SPD.

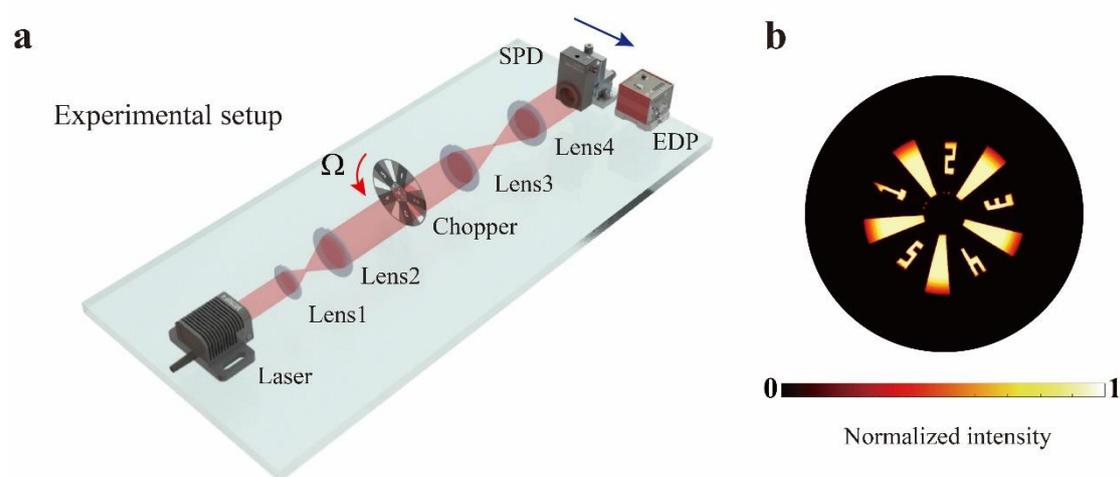

**Fig. 4 | Experimental configuration with SPD. a**, The experimental setup. The object is a chopper covered by numbers 1-5, which rotates at 100 Hz frequency. The SPD is scanned along the radial direction. **b**, The reconstructed image. SPD: single-point detector. EDP: electric displacement platform (see Fig. S7 in Supplementary Information for details).

We have proposed and experimentally demonstrated the use of LG transform for detecting a rotating object. Our method can have profound applications in scientific research and industry. For example, it provides an effective and powerful platform for in-situ monitoring of aero-engine, super-centrifuge, and gas turbine, which typically work under extreme conditions such as high pressure, high temperature, and high vacuum. Several unique advantages distinguish our method from traditional FT techniques in detecting a rotating object. First, the LG spectrum of a rotating object does not change, which provides a great convenience for real-time image reconstruction and analysis. Second, rotation itself is a feasible way to measure the LG spectrum. So, the experimental measurement system is simple and effective. Third, a commercial SPD can satisfy the detection requirement. In comparison to image cameras, the performance of low-cost SPDs are generally superior in detection efficiency, bandwidth, and response time. In addition, the demonstrated SPD configuration can be further extended to cover infrared, THz, and X-ray bands, where it is expensive or impossible to acquire a high-quality camera. Therefore, the proposed LG transform can truly provide a practical platform to detect and monitor high-speed rotating objects in biology, industry and astronomy.

**Methods**

**Image processing by manipulating the LG spectrum**

The azimuthal and radial edge enhancements (Figs. 3c and 3d in the main text) are achieved by choosing $\phi$ and $r$ as differential variables, respectively, i.e.,

$$\sum_l \sum_p A_{p,l} \left( \frac{\partial LG_{p,l}(r,\phi)}{r\partial \phi} \right) = \sum_l \sum_p i \frac{l}{r} A_{p,l} LG_{p,l}(r,\phi) = E_\phi, \quad (6)$$

$$\sum_l \sum_p A_{p,l} \left( \frac{\partial LG_{p,l}(r,\phi)}{\partial r} \right)$$
$$= \sum_l \sum_p \left[ \frac{|l|}{r} - \frac{2r}{\omega_0^2} - \frac{4r}{\omega_0^2} \frac{L_{p-1}^{|l|+1}\left(\frac{2r^2}{\omega_0^2}\right)}{L_p^{|l|}\left(\frac{2r^2}{\omega_0^2}\right)} \right] A_{p,l} LG_{p,l}(r,\phi) = E_r. \quad (7)$$

By combining Eqs. (6) and (7), we obtain the complete edge enhancement (Fig. 3e in the main text), i.e.,

$$\sum_l \sum_p A_{p,l} \left( \frac{\partial LG_{p,l}(r,\phi)}{\partial r} + \frac{\partial LG_{p,l}(r,\phi)}{r\partial \phi} \right)$$
$$= \sum_l \sum_p \left[ \frac{|l|}{r} - \frac{2r}{\omega_0^2} - \frac{4r}{\omega_0^2} \frac{L_{p-1}^{|l|+1}\left(\frac{2r^2}{\omega_0^2}\right)}{L_p^{|l|}\left(\frac{2r^2}{\omega_0^2}\right)} + i\frac{l}{r} \right] A_{p,l} LG_{p,l}(r,\phi). \quad (8)$$
$$= E_r + E_\phi = E_{r,\phi}$$

Here, $E_\phi$, $E_r$, and $E_{r,\phi}$ represent the obtained field after azimuthal, radial and entire edge enhancements, respectively.

The rotation operation (Fig. 3h in the main text) is achieved by simply adding a phase $l\Delta\phi$ to each LG mode, i.e.,

$$\begin{aligned}
&\sum_l \sum_p A_{p,l} LG_{p,l}(r,\phi) \exp(il\Delta\phi) \\
&= \sum_l \sum_p A_{p,l} LG_{p,l}(r) \exp(il\phi) \exp(il\Delta\phi) \\
&= \sum_l \sum_p A_{p,l} LG_{p,l}(r) \exp\left[il(\phi+\Delta\phi)\right] \\
&= E(r,\phi+\Delta\phi).
\end{aligned} \quad (9)$$

We use $\Delta\phi = 60°$ for example in Fig. 3h of the main text.

The reflection operation (Fig. 3i in the main text) is realized by conjugating the spiral phase of each LG component, which is described as

$$\sum_l \sum_p A_{p,l} \left[ LG_{p,l}(r,\phi) \right]^*$$
$$= \sum_l \sum_p A_{p,l} LG_{p,l}(r) \exp(-il\phi) \tag{10}$$
$$= E(r,-\phi)$$

By combining Eqs. (9) and (10) together, the image can be folded along an arbitrary axis. This operation can be described as

$$\sum_l \sum_p A_{p,l} \left[ LG_{p,l}(r,\phi) \right]^* \exp(il\Delta\phi)$$
$$= \sum_l \sum_p A_{p,l} LG_{p,l}(r) \exp\left[ il(-\phi + \Delta\phi) \right] \tag{11}$$
$$= E(r,-\phi + \Delta\phi)$$

The image is folded along $-\Delta\phi/2$. Figure 3j in the main text shows an example with $\Delta\phi = 60°$.

The pattern replication (Figs. 3m-o in the main text) is realized by superposing the LG modes with their $l$ indices being the multiple of an integer $m$, i.e.,

$$\sum_{l=km} \sum_p A_{p,l} LG_{p,l}(r,\phi) = E^m, \tag{12}$$

where $k = \pm 1, \pm 2, \pm 3 \cdots$ and $E^m$ is the obtained field after pattern replication operation.

**Data availability**

The data that supports the results within this paper and other findings of the study are available from the corresponding authors upon reasonable request.

**Code availability**

The custom code and mathematical algorithm used to obtain the results within this paper are available from the corresponding authors upon reasonable request.

**Acknowledgments**

We thank Jianping Ding for useful discussion. This work was supported by the National Key R&D Program of China (2017YFA0303703 and 2016YFA0302500), the National Natural Science Foundation of China (NSFC) (91950206 and 11874213), and Nanjing University Innovation and Creative Program for PhD candidate (CXCY17-27), Natural Science Foundation of Jiangsu Province (BK20180322) and the Fundamental Research Funds for the Central Universities (1480605201).

**Author contributions:** Y.Z. conceived the idea. D.W. and J.M. contributed equally to this work. D.W., J.M., T.W., C.X. performed the experiments and numerical simulations under the guidance of Y.Z., S.Z. and M.X. Y.C., L.Z., X.F., and D.Z.W. contributed to the discussion of experimental results. Y.Z. and M.X. supervised the project. D.W., J.M., Y.Z. and M.X. wrote the manuscript with contributions from all co-authors.

**Competing interests:** The authors declare no competing interests.

**Additional Information:**

**Supplementary information** is available for this paper.

**Correspondence and requests for materials** should be addressed to Y.Z.